\documentclass[11pt]{article}
\usepackage{subfig}
\usepackage{epsfig}
\usepackage{times}
\usepackage{amsfonts}
\usepackage{amsmath}
\usepackage{graphicx}
\usepackage{url}
\usepackage{amssymb}
\usepackage[text={6in,8.5in},centering]{geometry}
\usepackage{algorithm}
\usepackage{algorithmic}

\newcommand{\argmin}{\operatornamewithlimits{argmin}}
\title{Training a Large Scale Classifier with the \\
Quantum Adiabatic Algorithm}

\author{Hartmut Neven \\
{\it Google} \\
\texttt{neven@google.com} \\
\and
Vasil S. Denchev\\
{\it Purdue University}\\
\texttt{vdenchev@purdue.edu} \\
\and
\\
Geordie Rose and William G. Macready \\
{\it D-Wave Systems Inc.}\\
\texttt{\{rose,wgm\}@dwavesys.com}
}
\begin{document}
\maketitle
\algsetup{indent=2em}
\newcommand{\factorial}{\ensuremath{\mbox{\sc Factorial}}}
\newcommand{\bra}[1]{\langle #1|}
\newcommand{\ket}[1]{|#1\rangle}
\begin{abstract}
\par In a previous publication we proposed discrete global optimization as a method to train a strong
binary classifier constructed as a thresholded sum over weak classifiers. Our motivation was
to cast the training of a classifier into a format amenable to solution by the quantum adiabatic
algorithm. Applying adiabatic quantum computing (AQC) promises to yield solutions that are superior
to those which can be achieved with classical heuristic solvers. Interestingly we found that by using
heuristic solvers to obtain approximate solutions we could already gain an advantage over the
standard method AdaBoost. In this communication we generalize the baseline method to large scale classifier training.
By large scale we mean that either the cardinality of the dictionary of candidate weak classifiers or the number of weak learners
used in the strong classifier exceed the number of variables that can be handled effectively
in a single global optimization. For such situations we propose an iterative and piecewise approach
in which a subset of weak classifiers is selected in each iteration via global optimization.
The strong classifier is then constructed by concatenating the subsets of weak classifiers.
We show in numerical studies that the generalized method again successfully competes with AdaBoost.
We also provide theoretical arguments as to why the proposed optimization
method, which does not only minimize the empirical loss but also adds L0-norm regularization, is superior to versions of boosting that
only minimize the empirical loss. By conducting a Quantum Monte Carlo simulation we gather evidence that
the quantum adiabatic algorithm is able to handle a generic training problem efficiently.
\end{abstract}
\newpage
\section{Baseline System}
In \cite{Neven08b} we study a binary classifier of the form
\begin{equation} y = H(x)={\rm sign}\displaystyle \left(\sum_{i = 1}^{N} w_i h_i(x)\right) \hbox{ , }
\end{equation}
where $x \in \mathbb{R}^M$ are the input patterns to be classified, $y \in \{-1,+1\}$ is the output
of the classifier, the $h_i:x \mapsto \{-1, +1\}$ are so-called weak classifiers or features detectors,
and the $w_i \in \{0, 1\}$ are a set of weights to be optimized during training. $H(x)$ is known as a
strong classifier.

Training, i.e. the process of choosing the weights $w_i$, proceeds by simultaneously minimizing two
terms. The first term, called the loss $L(w)$, measures the error over a set of $S$ training examples
$\{(x_s, y_s) | s = 1, \ldots, S\}$. We choose least squares as the loss function. The second term, known
as regularization $R(w)$, ensures that the classifier does not become too complex. We employ a
regularization term based on the L0-norm, $\parallel w \parallel _0$. This term encourages the strong
classifier to be built with as few weak classifiers as possible while maintaining a low training error.
Thus, training is accomplished by solving the following discrete optimization problem:

\begin{eqnarray} w^{opt} = \arg\min_{w} \left(\underbrace{\sum_{s = 1}^{S} \left( {1\over N}\sum_{i = 1}^{N}
w_i h_i(x_s) - y_s \right)^2}_{L(w)} + \underbrace{\lambda \parallel w \parallel _0}_{R(w)} \right)\nonumber \\
= \arg\min_{w}\left({1\over N^2}\sum_{i = 1}^{N} \sum_{j = 1}^{N} w_i w_j \underbrace{\left(
\sum_{s = 1}^{S} h_i(x_s) h_j(x_s) \right)}_{Corr(h_i, h_j)} + \sum_{i = 1}^{N} w_i \left(\lambda - {2\over N}
\underbrace{\sum_{s = 1}^{S} h_i(x_s) y_s}_{Corr(h_i, y)}\right) \right)
\end{eqnarray}

\noindent Note that in our formulation, the weights are binary and not positive real numbers as in Ada\-Boost.
Even though discrete optimization could be applied to any bit depth representing the weights, we found
that a small bit depth is often sufficient \cite{Neven08b}. Here we only deal with the simplest case
in which the weights are chosen to be binary.

\section{Comparison of the baseline algorithm to AdaBoost}

\begin{figure}[b!]
\begin{center}
      \includegraphics[scale=1.05]{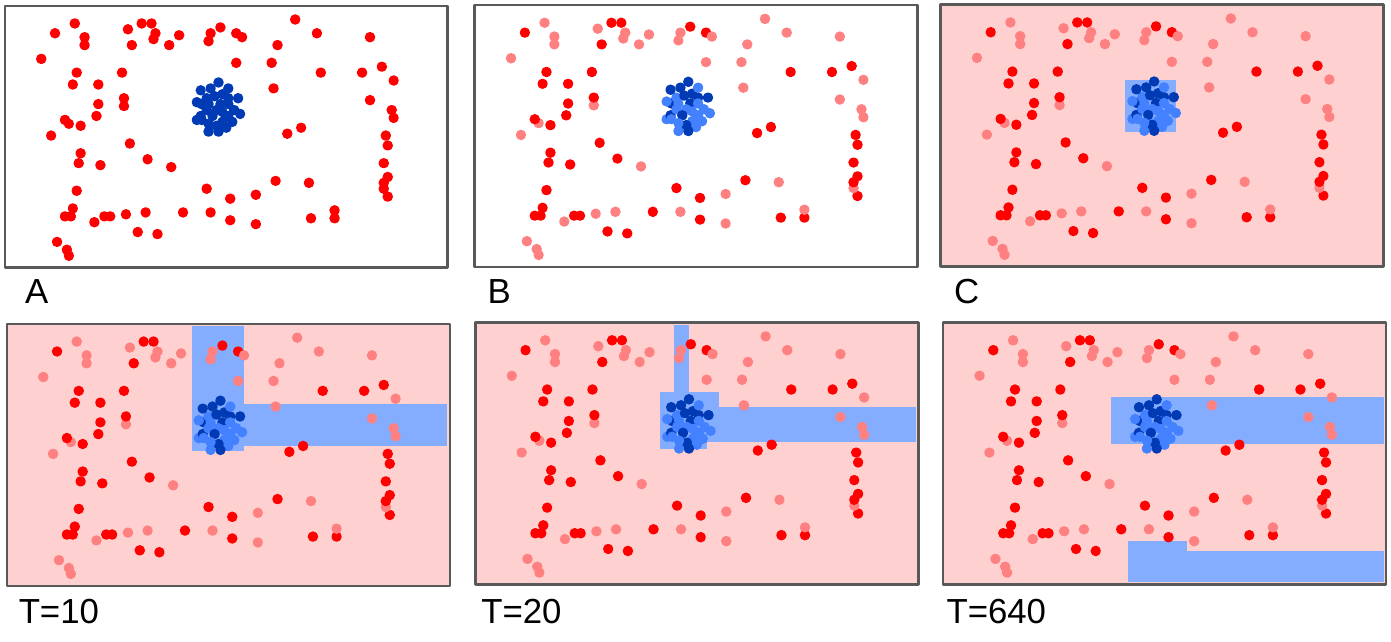}
      \caption{AdaBoost applied to a simple classification task. \textbf{A} shows the data, a separable set consisting of a
two-dimensional cluster of positive examples (blue) surrounded by negative ones (red). \textbf{B} shows the random
division into training (saturated colors) and test data (light colors). The dictionary of weak classifiers is constructed of axis-parallel
one-dimensional hyperplanes. \textbf{C} shows the optimal classifier for this situation, which
employs four weak classifiers to partition the input space into positive and a negative areas. The lower
row shows partitions generated by AdaBoost after 10, 20, and 640 iterations. The configuration at $T=640$
is the asymptotic configuration that does not change anymore in subsequent training rounds. The ''breakout regions"
outside the bounding box of the positive cluster occur in areas in which the training set
does not contain negative examples. This problem becomes more severe for higher dimensional data. Due to
AdaBoost's greedy approach, the optimal configuration is not found despite the fact that the
weak classifiers necessary to construct the ideal bounding box are generated. In fact AdaBoost fails to learn higher dimensional versions of this problem
altogether with error rates approaching 50\%. See section 6 for a discussion on how global optimization based learning can handle this data set.}
      \label{fig:synth1}
\end{center}
\end{figure}

In the case of a finite dictionary of weak classifiers $\{h_i(x)|i=1,...,N\}$ AdaBoost can be seen as a greedy
algorithm that minimizes the exponential loss \cite{Zhang2004},
\begin{equation}
\alpha^{opt} = \arg\min_{\alpha} \left(\sum_{s = 1}^{S} \exp\left( - y_s \sum_{i = 1}^{N}
\alpha_i h_i(x_s)\right)\right) \hbox{ , }
\end{equation}

\noindent with $\alpha_i \in \mathbb{R}^+$. There are two
differences between the objective of our algorithm (Eqn. 2) and the one employed by AdaBoost. The first is that we
added L0-norm regularization. Second, we employ a quadratic loss function, while Adaboost
works with the exponential loss.

\noindent It can easily be shown that including L0-norm regularization in the objective in Eqn. (2) leads to
improved generalization error as compared to using the quadratic loss only. The proof goes as follows.
An upper bound for the Vapnik-Chernovenkis dimension of a strong classifier $H$ of the form
$H(x)=\sum_{t=1}^{T} h_t(x)$ is given by

\begin{equation}
     VC_H=2(VC_{\{ h_i \}}+1)(T+1)\log_{2}( e(T+1)) \hbox{ , }
\end{equation}

\noindent where $VC_{\{ h_i \}}$ is the VC dimension of the dictionary of weak classifiers \cite{FreundShapire1995}.
The strong classifier's generalization error $Error_{test}$ has therefore an upper bound given by \cite{VapnikChervonenkis1971}

\begin{equation}
     Error_{test} \leq  Error_{train} + \sqrt{ {VC_H \ln({2S \over VC_H}+1) + ln({9 \over \delta})
\over S } } \hbox{ . }
\end{equation}

\noindent It is apparent that a more compact strong classifier that achieves a given training error
$Error_{train}$ with a smaller number $T$ of weak classifiers (hence, with a smaller VC dimension $VC_H$),
comes with a guarantee for a lower generalization error. Looking at the optimization problem in Eqn. 2,
one can see that if the regularization strength $\lambda$ is chosen weak enough, i.e.
$\lambda < {2 \over N}+ {1\over N^2}$, then the effect of the regularization is merely to thin out the
strong classifier. One arrives at the condition for $\lambda$ by demanding that the reduction of the
regularization term $\Delta R(w)$ that can be obtained by switching one $w_i$ to zero is smaller than the
smallest associated increase in the loss term $\Delta L(w)$ that comes from incorrectly labeling a
training example. This condition guarantees that weak classifiers are not eliminated at the expense of a
higher training error. Therefore the regularization will only keep a minimal set of components,
those which are needed to achieve the minimal training error that was obtained when using the loss term only.
In this regime, the VC bound of the resulting strong classifier is lower or equal to the VC bound of a
classifier trained with no regularization.

AdaBoost contains no explicit regularization term and it can easily happen that the classifier
uses a richer set of weak classifiers than needed to achieve the minimal training error, which in turn
leads to degraded generalization. Fig. 1 illustrates this fact.

In practice we do not operate in the weak $\lambda$ regime but rather determine the regularization
strength $\lambda$ by using a validation set. We measure the performance of the classifier for
different values of $\lambda$ on a validation set and then choose the one with the minimal validation
error. In this regime, the optimization performs a trade-off and accepts a higher empirical loss if the
classifier can be kept more compact. In other words it may choose to misclassify training examples if it can keep
the classifier simpler. This leads to increased robustness in the case of noisy data, and indeed we
observe the most significant gains over AdaBoost for noisy data sets when the Bayes error is high. The
fact that boosting in its standard formulation with convex loss functions and no regularization is not
robust against label noise has drawn attention recently \cite{LongRocco2008}\cite{Freund2009}.

The second difference to our baseline system, namely that we employ quadratic loss while AdaBoost works
with exponential loss is of smaller importance. In fact, the discussion above about the role of the
regularization term would not have changed if we were to choose exponential rather than square loss.
Literature seems to agree that the use of exponential loss in AdaBoost is not essential and that other
loss functions could be employed to yield classifiers with similar performance
\cite{Friedman1998}\cite{Wyner2002}. From a statistical perspective, quadratic loss is
satisfying since a classifier that minimizes the quadratic loss is consistent. With an increasing number of
training samples it will asymptotically approach a Bayes classifier i.e. the
classifier with the smallest possible error \cite{Zhang2004}.

\section{Generalization to large scale classifiers}
\par
The baseline system assumes a fixed dictionary containing a number of weak classifiers small enough, so
that all weight variables can be considered in a single global optimization. This approach needs to be
modified if the goal is to train a large scale classifier. Large scale here means that at least one of
two conditions is fulfilled:
\begin{enumerate}
    \item The dictionary contains more weak classifiers than can be considered in a single global
optimization.
    \item The final strong classifier consists of a number of weak classifiers that exceeds the number
of variables that can be handled in a single global optimization.
\end{enumerate}
\noindent Let us take a look at typical problem sizes. The state-of-art heuristic solver ILOG CPLEX can obtain good solutions for up to 1000 variable quadratic binary programmes depending on the coefficient matrix.
The quantum hardware solvers manufactured by D-Wave currently can handle 128 variable problems.
In order to train a strong classifier we often sift through millions of features. Moreover, dictionaries of weak learners are often dependent on a set of
continuous parameters such as thresholds, which means that their cardinality is infinite. We estimate that typical
classifiers employed in vision based products today use thousands of weak learners. Therefore
it is not possible to determine all weights in a single global optimization, but rather it is necessary
to break the problem into smaller chunks.

Let $T$ designate the size of the final strong classifier and $Q$ the number of variables that we can
handle in a single optimization. $Q$ may be determined by the number of available qubits, or if we employ
classical heuristic solvers such as ILOG CPLEX or tabu search \cite{Palubeckis2004}, then $Q$ designates a problem size for which we
can hope to obtain a solution of reasonable quality. We are going to consider two cases. The first with
$T \leq Q$ and the second with $T > Q$.

We first describe the "inner loop" algorithm we suggest if the number of variables we can handle exceeds
the number of weak learners needed to construct the strong classifier.

\begin{algorithm}
\caption{$T \leq Q$ (Inner Loop)}\label{alg:innerLoop}
\begin{algorithmic}[1]
\REQUIRE Training and validation data, dictionary of weak classifiers
\ENSURE Strong classifier
\medskip
\small{
	\STATE Initialize weight distribution $d_{inner}$ over training samples as
	uniform distribution $\forall s: d_{inner}(s) ={1\over S} \hspace{0.3cm}$
	\STATE Set $T_{inner}=0$
    \REPEAT {
	    \STATE Select the $Q-T_{inner}$ weak classifiers $h_i$ from the dictionary
	    that have the smallest weighted training error rates
	    \FOR{$\lambda =\lambda_{min}$ to $\lambda_{max}$}
			\STATE Run the optimization $w^{opt} = \arg\min_{w} \left(\sum_{s = 1}^{S}
			( {1\over Q}\sum_{i = 1}^{Q} w_i h_i(x_s) - y_s )^2 + \lambda \parallel w
			\parallel _0 \right)$
		    \STATE Set $T_{inner}=\parallel w \parallel _0$
		    \STATE Construct strong classifier $H(x)={\rm sign}\left(\sum_{t = 1}^{T_{inner}} h_t(x)\right)$
		    summing up the weak classifiers for which $w_i=1$
		    \STATE Measure validation error $Error_{val}$ of strong classifier on unweighted
		    validation set
		\ENDFOR
	    \STATE Keep $T_{inner}$, $H(x)$ and $Error_{val}$ for the $\lambda$ run
	    that yielded the smallest validation error
	    \STATE Update weights $d_{inner}(s)=d_{inner}(s)\left({1\over T_{inner}}\sum_{t = 1}^{T_{inner}} h_t(x) - y_s\right)^2$
	    \STATE Normalize $d_{inner}(s)={d_{inner}(s) \over
	    {\sum_{s=1}^{S}d_{inner}(s)}}$ }
    \UNTIL {validation error $Error_{val}$ stops decreasing}
}
\end{algorithmic}
\end{algorithm}

\noindent A way to think about this algorithm is to see it as an enrichment process. In the first round, the algorithm selects those $T_{inner}$ weak classifiers out of subset of Q that produce the optimal validation error. The subset of Q weak classifiers has been preselected from a dictionary with a cardinality possibly much larger than Q. In the next step the algorithm fills the $Q-T_{inner}$ empty slots in the solver with the best weak classifiers drawn from a modified dictionary that was adapted by taking into account for which samples the strong classifier constructed in the first round is already good and where it still makes errors. This is the boosting idea. Under the assumption that the solver always finds the global minimum, it is guaranteed that for a given $\lambda$ the solutions found in the subsequent round will have lower or equal objective value i.e. they achieve a lower loss or they represent a more compact strong classifier. The fact that the algorithm always considers groups of $Q$ weak classifiers simultaneously rather than incrementing the strong classifier one by one and then tries to find the smallest subset that still produces a low training error allows it to find optimal configurations more efficiently.

If the validation error cannot be decreased any further using the inner loop, one may conclude that more weak classifiers are needed to construct the strong one. In this case the "outer loop" algorithm "freezes" the classifier obtained so far and adds another partial classifier trained again by the inner loop.

\vspace{2mm}

\begin{algorithm}[h!]
\caption{$T > Q$ (Outer Loop)}\label{alg:outerLoop}
\begin{algorithmic}[1]
\REQUIRE Training and validation data, dictionary of weak classifiers
\ENSURE Strong classifier
\medskip
\small{
	\STATE Initialize weight distribution $d_{outer}$ over training samples as
	uniform distribution $\forall s: d_{outer}(s) ={1\over S} \hspace{0.3cm}$
    \STATE Set $T_{outer}=0$
    \REPEAT {
         \STATE Run \begin{bf}Algorithm 1\end{bf} with $d_{inner}$
         initialized from current $d_{outer}$ and using an
         objective function that takes into account the current $H(x)$:\\ $w^{opt} =
         \arg\min_{w} \left(\sum_{s = 1}^{S} ( {1\over {T_{outer}+Q}}
         (\sum_{t=1}^{T_{outer}} h_t(x_s)+ \sum_{i =
         1}^{Q} w_i h_i(x_s)) - y_s )^2 + \lambda \parallel w \parallel _0 \right)$
         \STATE Construct a strong classifier $H(x)={\rm sign}\left(\sum_{t=1}^{T_{outer}}
         h_t(x)+\sum_{t = T_{outer}+1}^{T_{outer}+T_{inner}} h_t(x)\right)$ adding those weak
         classifiers for which $w_i=1$
         \STATE Set $T_{outer}=T_{outer}+T_{inner}$
         \STATE Update weights
         $d_{outer}(s)=d_{outer}(s)\left({1\over T_{outer}}\sum_{t = 1}^{T_{outer}} h_t(x) - y_s\right)^2$
         \STATE Normalize $d_{outer}(s)={d_{outer}(s) \over {\sum_{s=1}^{S}d_{outer}(s)}}$ }
    \UNTIL {validation error $Error_{val}$ stops decreasing}
}
\end{algorithmic}
\end{algorithm}

\section{Implementation details and performance measurements}

\begin{figure}[t!]
\begin{center}
      \includegraphics[scale=0.65]{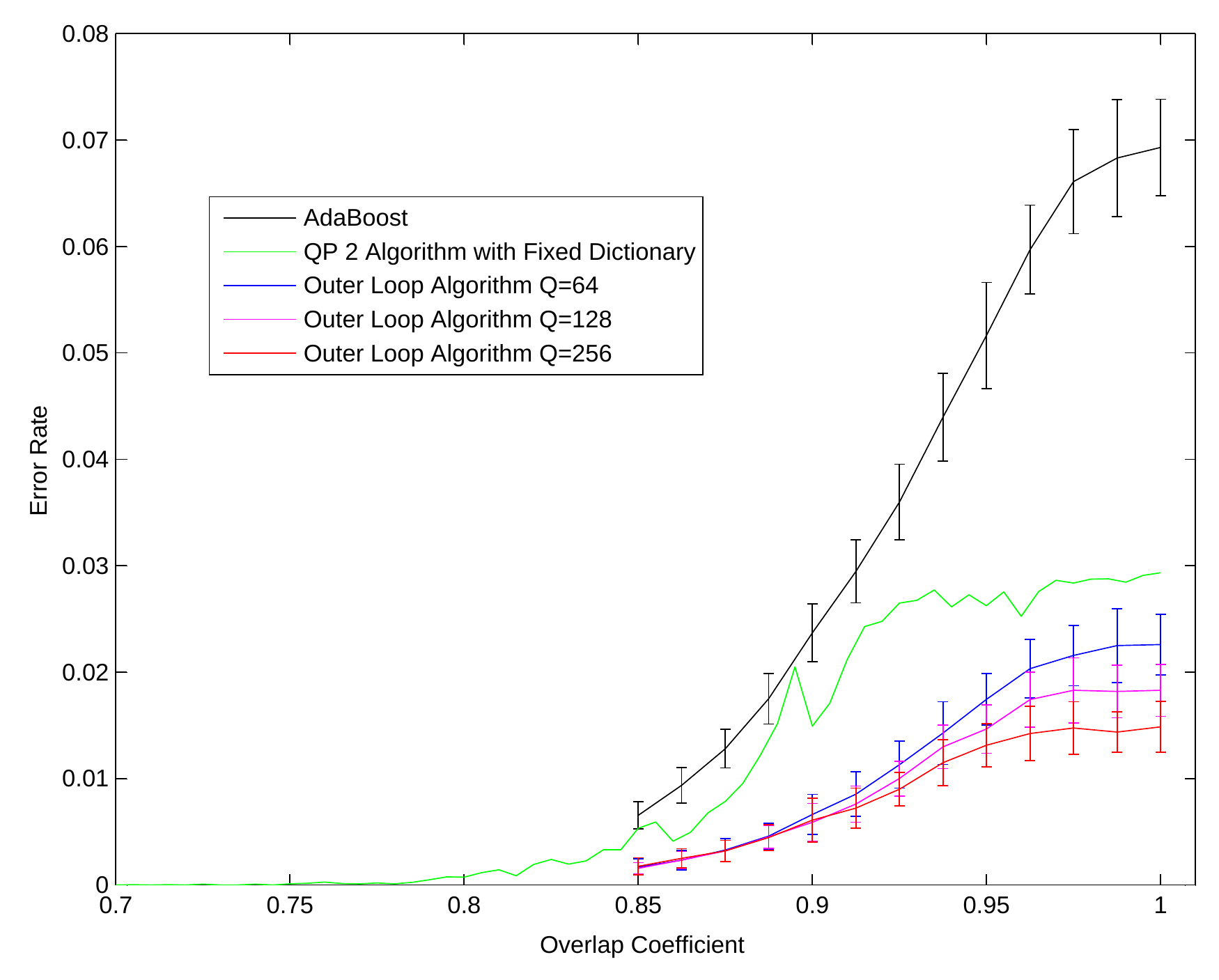}
      \caption{Test errors for the synthetic data set. We ran the outer loop algorithm for three different values of $Q$: $Q=64$, $Q=128$, and $Q=256$. The curves show means for 100 runs and the error bars indicate the corresponding standard deviations. All three versions outperform AdaBoost. The gain increases as the classification problem gets harder i.e. as the overlap between the positive and negative example clouds increases. The Bayes error rate for the case of complete overlap, overlap coefficient=1, is $\approx 0.05$. One can also see that there is a benefit to being able to run larger optimizations since the error rate decreases with increasing $Q$. For comparison, we also included the results from the last article \cite{Neven08b} for a classifier based on a fixed dictionary using quadratic loss (QP 2) for which the training was performed as per Eqn. 2. Not surprisingly, working with an adaptive set of weak classifiers yields higher accuracy.}
      \label{fig:synthetic}
\end{center}
\end{figure}

\begin{figure}[h!]
\begin{center}
      \includegraphics[scale=0.77]{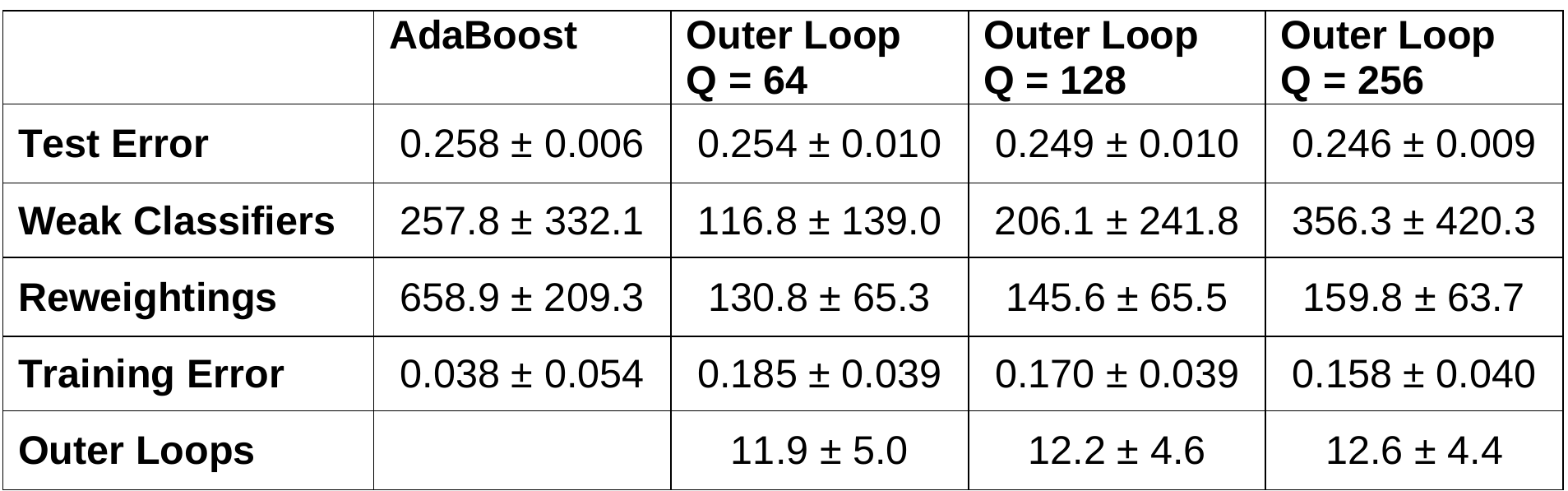}
      \caption{Test results obtained for the natural data set with 30-dimensional input vectors. Similar to the synthetic data we compare the outer loop algorithm for three different window sizes $Q=64$, $Q=128$, and $Q=256$ to AdaBoost. The results were obtained for 1000 runs. The piecewise global optimization based training only leads to sightly lower test errors but obtains those with a significantly reduced number of weak classifiers. Also, the number of iterations needed to train the strong classifiers is more than 4 times lower than required by AdaBoost.}
      \label{fig:natural1}
\end{center}
\end{figure}

\par To assess the performance of binary classifiers of the form (1) trained by applying the outer loop algorithm, we measured their performance on synthetic and natural data sets. The synthetic test data consisted of 30-dimensional input vectors generated by sampling from
$P(x,y) = {1\over2} \delta(y-1) N(x|\mu_+,I) + {1\over2} \delta(y+1) N(x|\mu_-,I)$, where
$N(x|\mu,\Sigma)$ is a spherical Gaussian having mean $\mu$ and covariance $\Sigma$. An
overlap coefficient determines the separation of the two Gaussians. See \cite{Neven08b} for details. The natural data consists of two sets of 30- and 96-dimensional vectors of Gabor
wavelet amplitudes extracted at eye locations in images showing faces. The input vectors are normalized using the L2-norm, i.e. we have
$\parallel x \parallel _2=1$. The data sets consisted of 20,000 input vectors, which
we divided evenly into a training set, a validation set to fix the parameter $\lambda$, and a test
set. We used Tabu search as the heuristic solver \cite{Palubeckis2004}. For both experiments we employed a dictionary consisting of decision stumps of the form:
\begin{equation} h_l^{1+}(x) = sign(x_l - \Theta_l^+)  \mbox{ for }  l = 1, \ldots, M \end{equation}
\begin{equation} h_l^{1-}(x) = sign(-x_l - \Theta_l^-)  \mbox{ for }  l = 1, \ldots, M \end{equation}
\begin{equation} h_l^{2+}(x) = sign(x_i x_j - \Theta_{i,j}^+)  \mbox{ for }   l = 1, \ldots, {M \choose 2}; i, j = 1, \ldots, M; i < j \end{equation}
\begin{equation} h_l^{2-}(x) = sign(-x_i x_j - \Theta_{i,j}^-) \mbox{ for }   l = 1, \ldots, {M \choose 2}; i, j = 1, \ldots, M; i < j \end{equation}
Here $h_l^{1+}$, $h_l^{1-}$, $h_l^{2+}$ and $h_l^{2-}$ are positive and negative
weak classifiers of orders 1 and 2 respectively; $M$ is the dimensionality of the input vector $x$; $x_{l}$,$x_{i}$,$x_{j}$ are the elements of the input vector and $\Theta_l^+$, $\Theta_l^-$, $\Theta_{i,j}^+$ and
$\Theta_{i,j}^-$ are optimal thresholds of the positive and negative weak classifiers of orders 1, and 2 respectively. For the 30-dimensional input data the dictionary employs 930 weak classifiers and for the 96-dimensional input it consists of 9312 weak learners.

As in \cite{Neven08b} we compute an optimal threshold for the final strong classifier according to
\hbox{$\Theta=\frac{1}{S}\sum_{s=1}^S \sum_{i=1}^{N} w_i^{opt}h_i(x_s)$}. The final classifier then becomes $y = {\rm sign} \left(\sum_{i = 1}^{N} w_i^{opt} h_i(x) - \Theta \right)$. In a separate set of experiments we co-optimized $\Theta$ jointly with the weights $w_i$. For the datasets we studied the difference was negligibly small but we do not expect this to be generally the case. Note that in order to handle the multi-valued global threshold within the frame work of discrete optimization one has to insert a binary expansion for $\Theta$ and the loss term then becomes $L(w)=\sum_{s = 1}^{S} \left( {1\over N}\left(\sum_{i = 1}^{N}
w_i h_i(x_s)-\sum_{k=0}^{\lceil \log_2 N \rceil}\Theta_{k}2^{k} +2^{\lceil \log_2 N \rceil}-1 \right) - y_s \right)^2$.


\par Test results for the synthetic data set are shown in Fig.~\ref{fig:synthetic} and the table in Fig.~\ref{fig:natural1} displays results obtained from the natural data set. We did comprehensive tests of the described inner and outer loop algorithms and found that minor modifications lead to the best results. We found that rather than adding just $Q-T_{inner}$ weak classifiers, the error rates dropped slightly (about 10\%) if we would replace all $Q$ classifiers from the previous round by new ones. The objective in Eqn. (2) employs a scaling factor of $1\over N$ to ensure that the unthresholded output of the classifier, sometimes referred to as score function, does not overshoot the $\{-1,+1\}$ labels. Systematic investigation of the scaling factor, however, suggested that larger scaling factors lead to a minimal improvement in accuracy and to a more significant reduction in the number of classifiers used (between 10-30\%). Thus, to obtain the reported results we chose a scale factor of $2\over {N}$.

To determine the optimal size $T$ of the strong classifier generated by AdaBoost we used a validation set. If the error did not decrease during 400 iterations we stopped and picked the T for which the minimal error was obtained. The results for the 96-dimensional natural data sets looked similar.

\section{Scaling analysis using Quantum Monte Carlo simulation}

\begin{figure} [b!]
\begin{center}
      \includegraphics[scale=0.7]{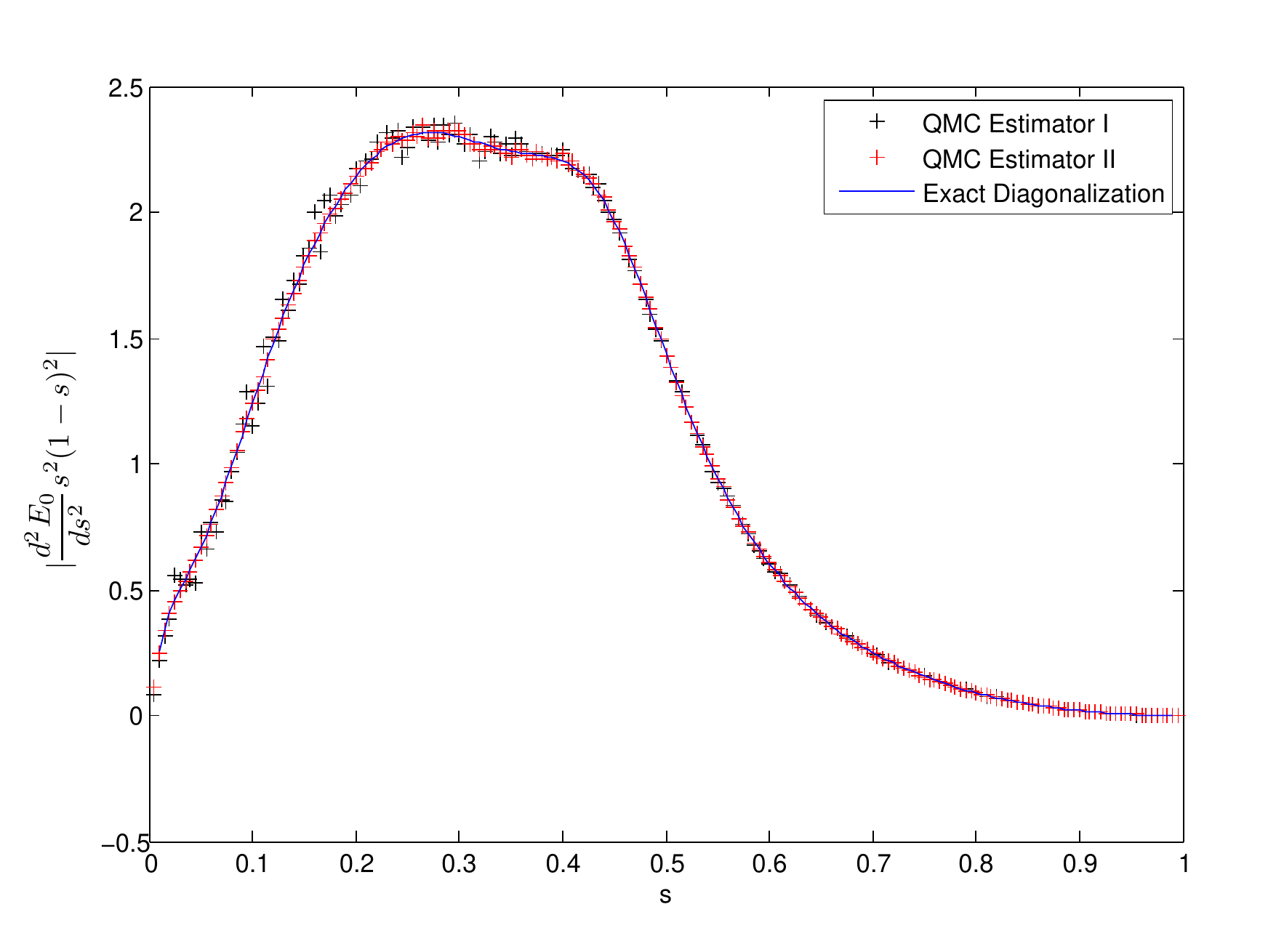}
      \caption{The quantity $|{d^2E_0\over ds^2} s^2 (1-s)^2|$
      determined by quantum Monte Carlo simulation as well as exact
      diagonalization for a training problem with 20 weight variables. For
      small problem instances with less than 30 variables, we can determine
      this quantity via exact diagonalization of the Hamiltonian $\tilde{H}(s)$.
      As one can see, the results obtained by diagonalization coincide very
      well with the ones determined by QMC. The training objective is given by
      Eqn. 2 using the synthetic data set with an overlap coefficient of 0.95.}
      \label{fig:QMC10}
\end{center}
\end{figure}

\begin{figure} [h!]
\begin{center}
      \includegraphics[scale=0.7]{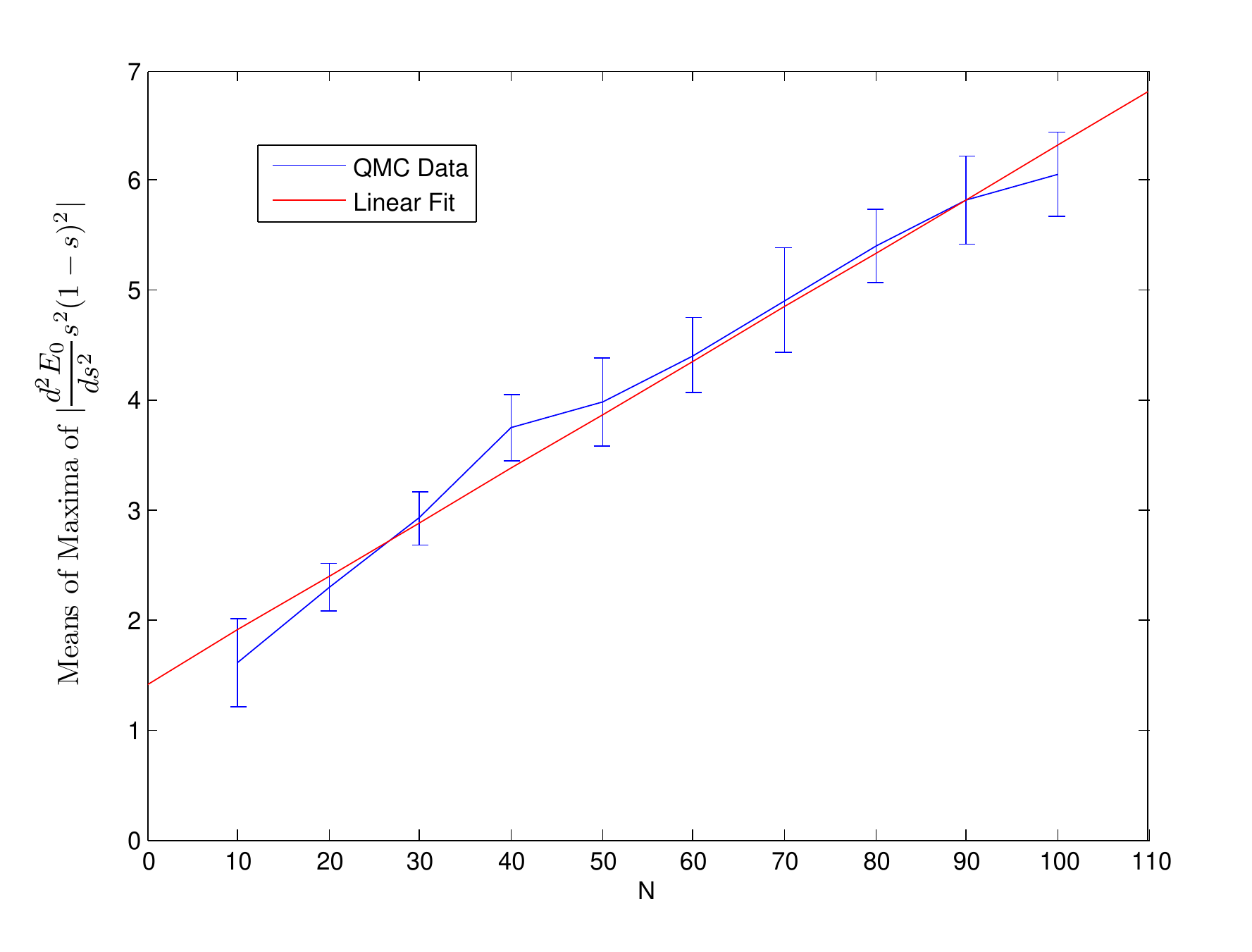}
      \caption{A plot of the peaks of the mean of $|{d^2E_0\over ds^2} s^2 (1-s)^2|$ against
      the number of qubits for the range of 10-100 qubits. The errors bars indicate the standard deviation. Each point of the plot represents 20 QMC runs.
      The data is well fitted by a linear function. From the fact that the scaling is at most
      polynomial in the problem size we can infer that the minimum gap and
      hence the runtime of AQC are scaling polynomially as well. \vspace{-0.3cm}}
      \label{fig:QMC20}
\end{center}
\end{figure}

We used the Quantum Monte Carlo (QMC) simulator of \cite{GossetFarhi2009} to
obtain an initial estimation of the time complexity of the
quantum adiabatic algorithm on our objective function. According to the adiabatic
theorem \cite{MessiahQMbook}, the ground state of the problem Hamiltonian $H_P$
is found with high probability by the quantum adiabatic algorithm, provided that
the evolution time $T$ from the initial Hamiltonian $H_B$ to $H_P$ is
$\Omega(g_{min}^{-2})$, where $g_{min}$ is the minimum gap. Here $H_B$ is
chosen as $H_B = \sum_{i=1}^N (1 - \sigma_i^x)/2$.
The minimum gap is the smallest energy gap between the ground state $E_0$
and first excited state $E_1$ of the time-dependent Hamiltonian $H(t) = (1 - t/T)H_B + (t/T)H_P$
at any $0 \leq t \leq T$. For notational convenience, we also use
$\tilde{H}(s) = (1 - s)H_B + sH_P$ with $0 \leq s \leq 1$. More
details can be found in the seminal work \cite{Farhi2000}.

\par As a consequence, to find the time complexity of AQC for a given objective
function, one needs to estimate the asymptotic scaling of the minimum gap
as observed on a collection of typical-case instances of this objective
function. As noted in \cite{AminChoi}, the task of analytically extracting the
minimum gap scaling has been extremely difficult in practice, except for a few
special cases. The only alternative is to resort to numerical methods, which
consist of diagonalization and QMC simulation. Unfortunately, diagonalization is
currently limited to about $N < 30$, and QMC to about $N=256$, where $N$ is the number
of binary variables \cite{Young2009}. Hence, the best that can be done with the
currently available tools is to collect data via QMC simulations on small problem instances and attempt to
extrapolate the scaling of the minimum gap for larger instances.

\par Using the QMC simulator of \cite{GossetFarhi2009}, we indirectly
measured the minimum gap by estimating the magnitude of the second
derivative of the ground state energy with respect to $s$, which is related to
the minimum gap \cite{Young2008}. This quantity is an upper bound on
$2|V_{01}|^2/g_{min}$, where $V_{01} = \bra{\Psi_0} d\tilde{H}/ds \ket{\Psi_1}$
and $\Psi_0, \Psi_1$ are the eigenstates corresponding to the ground and first
excited states of $\tilde{H}$. However, the quantity that one is interested in
for the time scaling of AQC is $|V_{01}|/g_{min}^2$, but assuming that the
matrix element $V_{01}$ is not extremely small, the scaling of the second
derivative, polynomial or exponential, can be used to infer whether the time
complexity of AQC is polynomial or exponential in $N$.

Fig. 5 shows the results of a scaling analysis for the synthetic data set. The
result is encouraging as the maxima of $|{d^2E_0\over ds^2}s^2(1-s)^2|$ only
scale linearly with the problem size. This implies that the runtime of AQC on this data set
is likely to scale at most polynomially. It is not possible to make a statement how
typical this data set and hence this scaling behavior is. We do know from
related experiments with known optimal solutions that Tabu search often fails
to obtain the optimal solution for a training problem for sizes as small as 64
variables\footnote{For instance we applied Tabu search to 30-dimensional
separable data sets of the form depicted in Fig.1. Tabu search failed to return
the minimal objective value for N=64 and S=9300.}. Obviously, scaling will
depend on the input data and the dictionary used. In fact it should be possible
to take a hard problem known to defeat AQC \cite{AminChoi} and encode it as a
training problem, which would cause scaling to become exponential. But even if
the scaling is exponential, the solutions found by AQC for a given problem size
can still be significantly better than those found by classical solvers.
Moreover, newer versions of AQC with changing paths \cite{GossetFarhi2009} may
be able to solve hard training problems like this efficiently.

\section{Discussion and future work}
\par Building on earlier work \cite{Neven08b} we continued our exploration of discrete global optimization as a tool to train binary classifiers. The proposed algorithms which we would like to call \emph{QBoost} enable us to handle training tasks of larger sizes as they occur in production systems today. QBoost offer gains over standard AdaBoost in three respects.

\begin{enumerate}
    \item The generalization error is lower.
    \item Classification is faster during execution because it employs a smaller number of weak classifiers.
    \item Training can be accomplished more quickly since the number of boosting
    steps is smaller.
\end{enumerate}

\noindent In all experiments we found that the classifier constructed with global optimization was significantly more compact. The gain in accuracy however was more varied.
The good performance of a large scale binary classifier trained using piecewise global optimization in a form amenable to AQC
but solved with classical heuristics shows that it is possible to map the training of a classifier to AQC with negative translation costs.
Any improvements to the solution of the training problem brought about by AQC will directly increase the advantage
of the algorithm proposed here over conventional greedy approaches. Access to emerging hardware that realizes the quantum adiabatic algorithm is needed
to establish the size of the gain over classical solvers. This gain will depend on the structure of the learning problem.

The proposed framework offers attractive avenues for extensions that we will explore in future work.
\\
\\
\noindent{\bf Alternative loss functions}
\\
\\
We employed the quadratic loss function in training because it maps naturally
to the quantum processors manufactured by D-Wave Systems which support solving
quadratic unconstrained binary programming. Other loss functions merit investigation as well including new versions that traditionally have not been studied by the machine learning community. A first candidate for which we have already done preliminary investigations is the 0-1 loss since it measures the categorical training error directly and not a convex relaxation. This loss function is usually discarded due to its computational intractability, making it an attractive candidate for the application of AQC. In particular 0-1 loss will do well on separable data sets with small Bayes error. An example is the dataset depicted in Fig. 1 and its higher-dimensional analogs. An objective as in Eqn. 2 employing 0-1 loss and including $\Theta$ in the optimization has the ideal solution as its minimum while for AdaBoost as well as the outer loop algorithm with square loss the error approaches 50\% as the dimension of the input grows larger.

We developed two alternative objective functions that mimic the action of the 0-1 loss in a quadratic optimization framework. Unfortunately this is only possible at the expense of auxiliary variables. The new objective minimizes the norm of the labels $\bar y_s$ simultaneously with finding the smallest set of weights $w_i$ that minimizes the training error. Samples that can not be classified correctly are flagged by error bits $e_s$.

\begin{eqnarray} &&({w}^{opt},{\bar y}^{opt}, e^{opt})= ... \nonumber\\
&=& \argmin_{w,\bar y, e} \Biggl( \sum_{s = 1}^{S} \Biggl( \left( \sum_{i = 1}^{N} w_i h_i(x_s) - sign(y_s){\bar y}_s \right)^2 + ...\nonumber \\
&&... + N^2 \left( \sum_{i = 1}^{N} w_i h_i(x_s) - sign(y_s){\bar y}_s + sign(y_s) N e_s \right)^2 \Biggr)+ \lambda \sum_{i = 1}^{N} {w_i} \Biggr) \nonumber \\
&=& \argmin_{w,\bar y, e} \Biggl(  (1+N^2) \sum_{i,j = 1}^{N} w_i w_j
	\left(\sum_{s = 1}^{S} h_i(x_s) h_j(x_s) \right) + ...\\
&&... + (1+N^2) \sum_{s = 1}^{S} (\bar {y}_{\dagger}^2 + 2 \bar
	{y}_{\dagger} \sum_{k=0}^{\lceil \log_2 N \rceil-1}\bar y_{k,s}2^{k} +
	\sum_{k,k^\prime = 0}^{\lceil \log_2 N \rceil-1} \bar y_{k,s}\bar
	y_{k^\prime,s}2^{(k+k^\prime)}) +... \nonumber\\
&&... + N^4 \sum_{s = 1}^{S} e_s - 2(1+N^2) \sum_{i = 1}^{N} \sum_{s = 1}^{S}
	w_i sign(y_s)h_i(x_s) (\bar {y}_{\dagger} + \sum_{k = 0}^{\lceil \log_2 N
	\rceil-1} \bar y_{k,s} 2^{k}) +  ...\nonumber\\
&&... + 2 N^3 \sum_{i = 1}^{N} \sum_{s = 1}^{S} w_i e_s sign(y_s)h_i(x_s) - 2
	N^3 \sum_{s = 1}^{S} e_{s}(\bar {y}_{\dagger} + \sum_{k = 0}^{\lceil \log_2 N
	\rceil-1} \bar y_{k,s} 2^{k}) + \lambda \sum_{i = 1}^{N} {w_i}\Biggr)\nonumber
\end{eqnarray}

\noindent with $\bar y_s \in \{1,2,,...,N\}$. To replace the N-valued ${\bar y}_s$ with binary variables
we effected a binary expansion
${\bar y}_s = \bar y_{\dagger} + \sum_{k = 0}^{\lceil \log_2 N \rceil-1}\bar
y_{k,s}2^k$.  $\bar {y}_{\dagger}$ is a constant we set to 1 for the purpose of preventing
${\bar y}_s$ from ever becoming 0. The number of binary variables needed is $N$ for $w$, $S \lceil \log_2 N \rceil$ for $\bar y$ and $S$ for $e$.

The computational hardness of learning objectives based on 0-1 loss manifested itself in that for handcrafted datasets for which we knew the solution we could see that Tabu search was not able to find the minimum assignment. We also conducted a QMC analysis but were not able to determine a finite gap size for problem sizes of 60 variables and larger. However this was a preliminary analysis that will have to be redone with larger computational resources. The difficulty to determine the gap size led us to propose an alternative version that uses a larger number of auxiliary variables but has a smaller range of coefficients. Samples that can be classified correctly are flagged by indicator bits $e_s^{+}$. Vice versa samples that can not be classified correctly are indicated by $e_s^{-}$.

\begin{eqnarray} &&({w}^{opt},{\bar y}^{opt}, e^{opt})= ... \\
&=& \argmin_{w,\bar y, e} \Biggl( \sum_{s = 1}^{S} \Biggl( \left( \sum_{i = 1}^{N} w_i h_i(x_s) - (e_s^{+} - e_s^{-}) sign(y_s){\bar y}_s \right)^2 +  e_s^{-}  \Biggr)+ \lambda \sum_{i = 1}^{N} {w_i} \Biggr)\nonumber
\end{eqnarray}

The number of binary variables needed is $N$ for $w$, $S \lceil \log_2 N
\rceil$ for $\bar y$ and $2S$ for the $e_s^{+}$ and $e_s^{-}$. However since
the objective contains third-order terms, we will need to effect a variable
change to reduce to second order: $y_s^{+}=e_s^{+} \bar y_s$ and
$y_s^{-}=e_s^{-} \bar y_s$. This adds another $2S\lceil \log_2 N \rceil$
qubits. Due to the large number of binary variables we have not analyzed
Eqn. 11 yet.
\\
\\
{\bf Co-Optimization of weak classifier parameters}
\\
\\
The weak classifiers depend on parameters such as the thresholds $\Theta_l$.
Rather than determining these parameters in a process outside of the global
optimization it would better keep with the spirit of our approach to include
these in the global optimization as well. Thus is would look more like a
perceptron but one in which all weights are determined by global optimization.
So far we have not been able to find a formulation that only uses quadratic
interactions between the variables and that does not need a tremendous amount of auxiliary variables. This is due to the fact that the weak classifier parameters live under the sign function which makes the resulting optimization problem contain terms of order $N$ if no simplifications are effected. Our desire to stay with quadratic optimization stems from the fact that the current D-Wave processors are designed to support this format, and that it will be hard to represent $N$-local interactions in any physical process.
\\
\\
{\bf Co-Training of multiple classifiers}
\\
\\
Our training framework allows for simultaneous training of multiple classifiers
with feature sharing in a very elegant way. For example if two classifiers are
to learn similar tasks, then a training objective is formed that sums up two
objectives as described in Eqn. (2)---one for each classifier. Then cross terms
are introduced that encourage the reuse of weak classifiers and which have the
form $\mu \sum_{i = 1}^{N} (w^A_i - w^B_i)^2$. The $w^A_i$ and $w^B_i$ are the
weights of classifiers A and B respectively. From the perspective of classifier
A this looks like a special form of context dependent regularization. The
resulting set of classifiers is likely to exhibit higher accuracy and reduced
execution times. But more importantly, this framework may allow reducing the
number of necessary training examples.
\\
\\
{\bf Incorporating Gestalt principles}
\\
\\
The approach also allows to seamlessly incorporate a priori knowledge about the
structure of the classification problem, for instance in the form of Gestalt
principles. For example, if the goal is to train an object detector, it may be
meaningful to impose a constraint that if a feature is detected at position $x$
in an image then there should also be one at a feature position $x'$ nearby.
Similarly, we may be able to express symmetry or continuity constraints by
introducing appropriate penalty functions on the weight variables optimized
during training. Formally, Gestalt principles take on the form of another
regularization term, i.e. a penalty term $G(w)$ that is a function of the weights.

\section*{Acknowledgments}

We would like to thank Alessandro Bissacco, Jiayong Zhang and Ulrich Buddemeier for their assistance with MapReduce, Boris Babenko for helpful discussions of approaches to boosting, Edward Farhi and David Gosset for their support with the Quantum Monte Carlo simulations, Corinna Cortes for reviewing our initial results and Hartwig Adam, Jiayong Zhang and Edward Farhi for reviewing drafts of the paper.

\bibliography{references}
\bibliographystyle{alpha}
\end{document}